\documentclass[apj]{emulateapj}
\usepackage{graphicx}
\usepackage{hyperref}
\usepackage{float}
\usepackage{amsmath}
\usepackage{color}

\slugcomment{as of \today}
\shorttitle{Mass Loss in 47 Tuc}
\shorttitle{Heyl et al.}

\begin{document}

\title{When do stars in 47 Tucanae lose their mass?}
\author{Jeremy~Heyl\altaffilmark{1}, Jason Kalirai\altaffilmark{2}, Harvey~B.~Richer\altaffilmark{1},
  Paola Marigo\altaffilmark{3},
Elisa Antolini\altaffilmark{4}, Ryan Goldsbury\altaffilmark{1},  Javiera Parada\altaffilmark{1} }
\altaffiltext{1}{Department of Physics \& Astronomy, University of
  British Columbia, Vancouver, BC, Canada V6T 1Z1;
  heyl@phas.ubc.ca, richer@astro.ubc.ca  }
\altaffiltext{2}{Space Telescope Science Institute,Baltimore MD
  21218; Center for Astrophysical Sciences, Johns Hopkins University, Baltimore MD, 21218; jkalirai@stsci.edu }
\altaffiltext{3}{Dipartimento di Fisica e Astronomia ``G. Galilei'', Universit\`a degli Studi di Padova, I-35122 Padova, Italia}
\altaffiltext{4}{Dipartimento di Fisica e Geologia, Universit\`a degli Studi di Perugia, I-06123 Perugia, Italia}

\begin{abstract}
  By examining the diffusion of young white dwarfs through the core of
  the globular cluster 47~Tucanae, we estimate the time when the
  progenitor star lost the bulk of its mass to become a white dwarf.
  According to stellar evolution models of the white-dwarf progenitors
  in 47 Tucanae, we find this epoch to coincide approximately with the
  star ascending the asymptotic giant branch ($3.0 \pm 8.1$~Myr before
  the tip of the AGB) and more than ninety million years after the
  helium flash (with ninety-percent confidence).  From the diffusion
  of the young white dwarfs we can exclude the hypothesis that the
  bulk of the mass loss occurs on the red-giant branch at the
  four-sigma level.  Furthermore, we find that the radial distribution
  of horizontal branch stars is consistent with that of the red-giant
  stars and upper-main-sequence stars and inconsistent with the loss
  of more than 0.2 solar masses on the red-giant branch at the
  six-sigma level.
\end{abstract}


%
\section{Introduction}
\label{sec:intro}

When and how a star like the Sun loses its mass to become a white
dwarf star is a key open question of stellar evolution.  Does the bulk
of the mass loss occur when the star is a red giant or when the star
reaches the asymptotic giant branch?  For forty years, in stellar
evolution models mass loss on the red-giant branch (RGB) has been
commonly described with the \citet{1975MSRSL...8..369R} formula, an
empirical scaling relation that involves basic stellar parameters and
an adjustable efficiency parameter, $\eta_{\rm RGB}$. This parameter
is usually constrained from the requirement of reproducing the
extended blue horizontal branches (HB) in the Hertzsprung-Russell
diagrams of Galactic globular clusters \citep{1988ARA&A..26..199R}.
Typical values are $\eta_{\rm RGB}\simeq 0.35-0.45$, which implies
that $\simeq 0.20-0.25 \, M_{\odot}$ is the mass that should be lost
on the red-giant branch by a star with initial mass $0.85-0.90\,
M_{\odot}$, characteristic of the turn-offs in Galactic globular
clusters. An additional mass loss event of smaller size, $\simeq
0.10-0.15 \, M_{\odot}$, should take place later, during the evolution
on the asymptotic giant branch (AGB).  Recently
\citet{2015MNRAS.448..502M} have also argued for values of $\eta_{\rm
  GB} = 0.477 \pm 0.070 ^{0.050}_{-0.062}$ for Galactic clusters and
$\eta_{\rm GB} = 0.452$ for 47 Tuc in particular.

On the other hand this classical paradigm has been recently challenged
from various perspectives.  Perhaps, one of the most intriguing
findings is that the extended blue horizontal branch may be well
explained by the very high helium abundance associated with the bluest
main sequence of the multiple populations widely present in Galactic
globular clusters \citep[e.g.,][]{Lee_etal05, DantonaCaloi_08}.  In general, stellar
models indicate that at the same age, higher helium corresponds to
lower turn-off mass \citep{Bertelli_etal08}.  Therefore, for the same
mass ejected on the red-giant branch, a higher helium abundance
favors the development of more extended horizontal branches.

At the same time there are hints that the mass loss rates predicted
for red-giant-branch stars with the classical calibrations may be
significantly overestimated.  Using chromospheric model calculations
of the H$\alpha$ line for a sample of red-giant-branch stars in a few
globular clusters \citet{Meszaros_etal09} pointed out that the
resulting mass loss rates are about one order of magnitude lower than
obtained with the Reimers law or inferred from the infrared excess of
similar stars by \citet[][see also \citet{Origlia_etal14} for recent
  results]{Origlia_etal07}.  Such discrepancy appears to be
overcome according to \citet{Groenewegen_12}, who found agreement
between the mass loss rates derived from fitting the spectral energy
distributions of red-giant-branch stars with infrared excess, and the
chromospheric estimates.  More recently
\citet{Groenewegen_14} has performed the first detection of rotational CO
line emission in a nearby red giant branch with a luminosity of
$\simeq 1300\, L_{\odot}$ and an estimated mass-loss rate as low as a
few $10^{-9}\,M_{\odot}$ yr$^{-1}$.  Interestingly
\citet{2012MNRAS.419.2077M} argued from Kepler asteroseismic
measurements of the stars in the very metal-rich open cluster NGC~6971
that low values of $\eta$ on the red-giant branch
($\eta_\mathrm{RGB}$) are needed to account for the mass loss between
the red giant and red clump phases of stars in this metal-rich cluster
($0.1\lesssim \eta_\mathrm{RGB} \lesssim 0.3$).

On the other hand, mass loss on the asymptotic giant branch of old
stellar populations may have been underestimated at lower
metallicities.  Recent stellar population synthesis studies have shown
that to reproduce the star counts and luminosity functions of
metal-poor low-mass thermally-pulsing asymptotic-giant branch (TP-AGB)
stars in a sample of nearby galaxies one has to invoke a more
efficient mass loss than the classical Reimers recipe \citep[][see
  Sect.~\ref{sec:stell-evol-models}] {Girardi_etal10,
  Rosenfield_etal14}. This also yields good agreement with the
low-mass end of the initial-final mass relation, as probed with the
white dwarfs in M\,4 \citep{2009ApJ...705..408K}.

\begin{figure*}
\includegraphics[width=\textwidth,clip,trim=80px 10px 80px 20px]{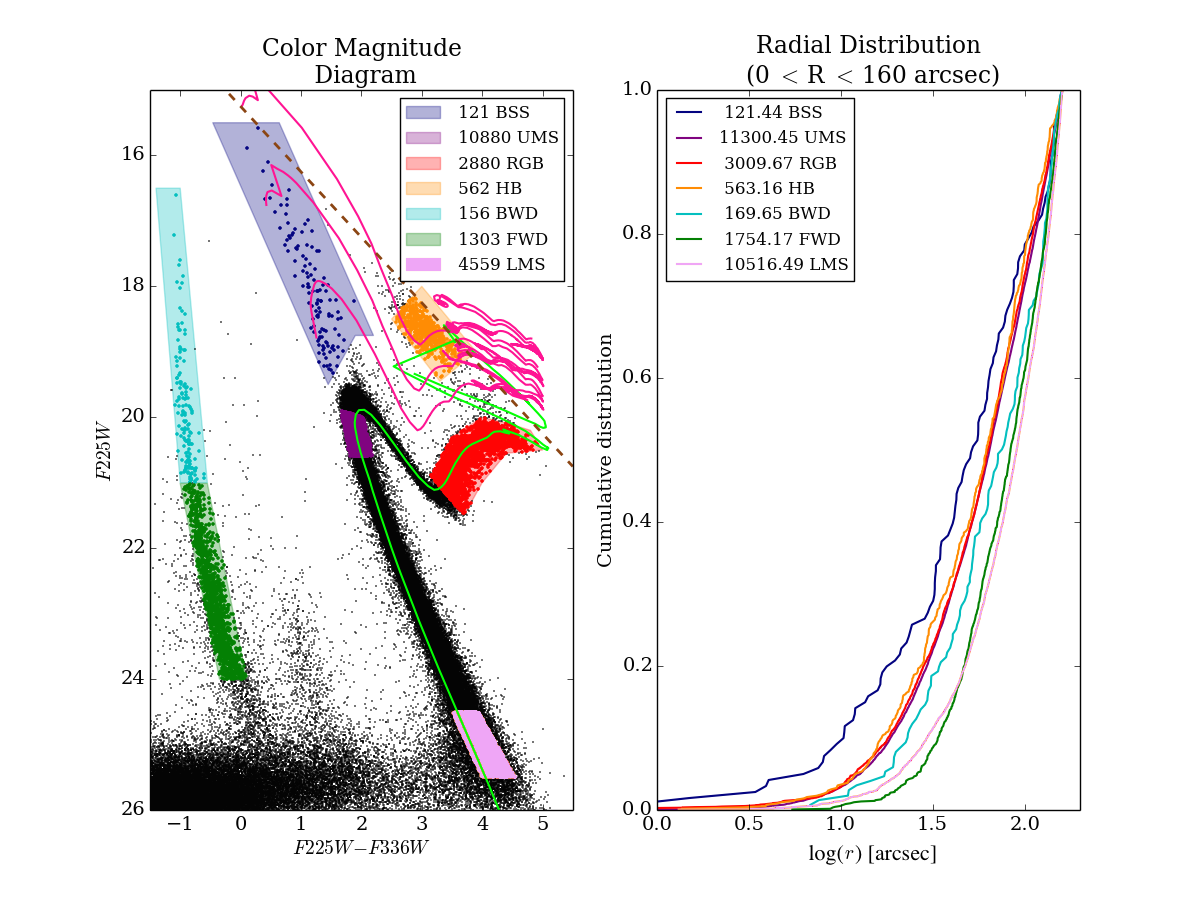}
\caption{Left: Color-magnitude diagram of the core of 47 Tucanae in
  the WFC3 filters F225W and F336W.  We have imposed a mild cut on
  magnitude error to remove strong saturated stars from the sample.
  The inset gives the number of stars in each class before correcting
  for completeness.  The classes are: BSS, blue straggler stars; UMS,
  upper-main-sequence stars of about $0.9\mathrm{M}_\odot$; RGB,
  red-giant-branch stars; HB, horizontal-branch stars; BWD, bright
  white dwarfs; FWD, faint white dwarfs; and LMS, lower-main-sequence
  stars of about $0.65\mathrm{M}_\odot$.  We determine locations of
  these classes on the color-magnitude diagram from the PARSEC
  isochrones and the MESA evolutionary models.  The magenta curves
  trace evolutionary tracks calculated with MESA and the atmosphere
  models of \citet{2004astro.ph..5087C}.  The magenta curves from
  bottom to top are for initial masses of 1.1, 1.4 and 1.8 solar
  masses.  We have assumed that the post-main sequence of the
  blue-straggler stars follows a similar track to normal stars of a
  similar mass \citep{2009ApJ...692.1411S}.  The green curve is a
  PARSEC isochrone\footnote{{\tt
      http://stev.oapd.inaf.it/cgi-bin/cmd\_2.7}} for an age of 11~Gyr
  and the metallacity of 47~Tucanae using the bolometric corrections
  of \citet{2014MNRAS.444.2525C}. An additional 0.4 and 0.3 magnitudes
  of extinction were added in F225W and F336W respectively to fit the
  isochrone to the data.  The dashed line traces $F336W=15.25$, the
  approximate magnitude where the images are saturated, so model
  curves to the right of this line will have observed colors just to
  the left of the line.  Right: The completeness-corrected radial
  distribution of stars in the various regions highlighted in the
  color-magnitude diagram.  The radial distribution of the upper-main
  sequence stars (UMS) and the red giants (RGB) are nearly
  indistinguishable. Here the inset gives the size of each sample
  after correcting for completeness.}
%
\label{fig:CMD}
\end{figure*}

In this study we build upon the results of \citet{Heyl14diff} to
demonstrate that the bulk of the mass loss from stars in 47 Tucanae
must happen on the asymptotic giant branch.  In the core of the
globular cluster 47~Tucanae the timescale for dynamical relaxation
through two-body interactions is similar to the stellar evolution
timescale for a star to live as a horizontal-branch star, rise up the
asymptotic branch and become a white-dwarf star
\citep{1996AJ....112.1487H,Heyl14diff}.  The core of 47~Tuc has been
the focus of numerous previous investigations
\citep[e.g][]{2006ApJS..166..249M, 2008ApJ...683.1006K,
  2009AJ....138.1455B, Milone_etal12}, but \citet{Heyl14diff} is the
first paper that combines the near ultraviolet filters of the Hubble
Space Telescope (HST) with a mosaic that covers the entire core of the
cluster.  In these filters the young white dwarfs, the giant stars,
the horizontal-branch stars, the blue stragglers and the upper main
sequence stars all have similar magnitudes as shown in
Fig.~\ref{fig:CMD}.  Furthermore, the exquisite angular resolution in
the near ultraviolet of the new Hubble WFC3/UVIS camera also reduces
the effects of confusion and incompleteness in this crowded field.
From theoretical arguments it has long been argued that two-body
interactions will sort the stars in a globular cluster by mass with
the more massive stars lying closer to the center of the cluster
\citep[e.g.][]{Spit87}, and this mass segregation has been
quantified in various clusters \cite[e.g][]{2013arXiv1308.3706G}.
\citet{Heyl14diff} for the first time caught this process of mass
segregation in action and determined the timescale for the sorting of
stars by mass, the relaxation time, to be about 30~Myr.  They
have outlined in detail how to measure the completeness rate and model
the diffusion of stars in the cluster using the young white dwarfs.

Here we build upon this diffusion model by comparing the radial
distribution of the white dwarfs with that of the upper-main-sequence
and red-giant stars.  The key observation that one can draw from
Fig.~\ref{fig:CMD} is that the distribution of the bright white
dwarfs, whose median age along the white-dwarf cooling track is 6~Myr,
more closely resembles that of the upper-main-sequence stars than that
of the fainter white dwarfs of about 100~Myr.  If the progenitors of
the white dwarfs lost their mass more than 30~Myr before the birth of
the white dwarfs, the white dwarfs would have already been sorted by
mass, so their radial distribution would not look so similar to that
of the upper-main sequence stars.  Furthermore, because the horizontal
branch in 47~Tucanae is thought to last for about 80~Myr, a few
relaxation times, their radial distribution will also reflect the mass
of the horizontal branch stars.  The radial distribution of the
horizontal branch stars is very similar to that of the
upper-main-sequence stars, their progenitors. We will confront these
observations with expectations from stellar evolution models and
better quantify it with the diffusion models from \citet{Heyl14diff}.

\section{Stellar Evolution Models}
\label{sec:stell-evol-models}

To construct the stellar evolution models here, we used both MESA
(Modules for Experiments in Stellar Astrophysics;
\citealt{2011ApJS..192....3P}) and a combination of PARSEC (for the
evolution before the TP-AGB; \citealt{Bressan2012}) and COLIBRI (for
the TP-AGB; \citealt{Marigo2013}) to perform simulations of stellar
evolution starting with a pre-main-sequence model of 0.9 solar masses
and a metallicity of $Z=4\times 10^{-3}$ and $Y=0.256$, appropriate
for the cluster 47~Tucanae \citep{2009AJ....138.1455B}, assuming
    $[\mathrm{Fe/H}]=-0.83$ and
    $[\alpha/\mathrm{Fe}]=0.3$.  We note that
adopting $[\mathrm{Fe/H}]=-0.76$ and individual elemental abundances
measured in 47 Tuc stars \citep[as summarized in ][]{Milone_etal12}
leads to a somewhat larger metallicity ($Z\sim0.0055$), but the
general trends, discussed below, do not change significantly.

For the MESA models, we used SVN revision 5456 and started with the
model {\tt 1M\_pre\_ms\_to\_wd} in the test suite.  We changed the
parameters {\tt initial\_mass} and {\tt initial\_z} of the star and
adjusted the parameter {\tt log\_L\_lower\_limit} to $-6$ so the
simulations run well into the white dwarf cooling regime.  We
also reduced the values of the wind $\eta$ on the RGB and AGB to 0.46
(from the default of 0.7) to yield a 0.53 solar mass white dwarf
\citep{1988ARA&A..26..199R,2004A&A...420..515M,2009ApJ...705..408K}
from the 0.9 solar mass progenitor.  The MESA models are consistent
with the observed $M_V$ of the tip of the TP-AGB
\citep{2005A&A...441.1117L} which is sensitive to the mass of the
resulting WD and with the observed cooling curve of the white dwarfs
\citep{Heyl14diff}. For the mass loss on the red-giant branch we used
the \citet{1975MSRSL...8..369R} value,
\begin{equation}
{\dot M}_R =   4 \times 10^{-13} \eta  \frac{L}{L_\odot} \frac{R}{R_\odot}
\frac{M_\odot}{M} \,\,\,\,\,\,\,\,\,\,\,\,\,[{M_\odot}{\mathrm{yr}}^{-1}]
\label{eq:1}
\end{equation}
and on the asymptotic-giant branch we use the \citet{1995A&A...297..727B} formula
\begin{equation}
{\dot M}_B = 4.83 \times 10^{-9} \left (\frac{M}{M_\odot} \right)^{-2.1} \left (\frac{L}{L_\odot} \right)^{2.7} {\dot M}_R.
\label{eq:2}
\end{equation}
These parameters yield a model where the star loses about 0.2 solar
masses as a red giant and 0.17 solar masses as an asymptotic giant
star.  As other options we also used a value of $\eta$ on the
red-giant branch of 0.1, 0.2 and 0.3 in the range of
\cite{2012MNRAS.419.2077M} and higher values on the AGB
($\eta_\mathrm{AGB}$) as outlined in Table~\ref{tab:windparam}.  These
also yielded a 0.53 solar mass white dwarf but with much less mass
loss on the red-giant branch with values of $\eta$ in better
accordance with the results of \citet{2012MNRAS.419.2077M}.

Table~\ref{tab:windparam} summarizes the results of the various wind
models using MESA.  Essentially the two wind parameters
$\eta_\mathrm{RGB}$ and $\eta_\mathrm{AGB}$ can be tuned to change the
ratio of mass loss on the two giant branches without changing the
initial or final mass of the star (here $0.9$ and
$0.53\mathrm{M}_\odot$), the age of the star where it becomes a white
dwarf at the tip of the AGB ($t_\mathrm{WD}$) --- this should equal
the age of the globular cluster today.  We examine two other
timescales.  The first is the time interval between the tip of the
red-giant branch and the tip of the asymptotic giant branch, $\Delta
t_\mathrm{T2T}$, and the second is the time between the helium flash
and when the central helium abundance drops below $10^{-5}$, $\Delta
t_\mathrm{CHB}$, the epoch of core helium burning.  During the core
helium burning stage of the star, the luminosity remains nearly
constant for about 80~Myr, somewhat less than these two timescales,
according to the PARSEC evolutionary tracks.  This is the expected
time that stars will linger within the region of the CMD denoted as
the horizontal branch in Fig.~\ref{fig:CMD}.  When the two wind
parameters are equal, the mass loss on the two branches also ends up
being about equal and using the range of parameters outlined by
\citet{2012MNRAS.419.2077M} one can have as little as one-ninth of the
mass loss on the red-giant branch, leaving nearly ninety percent of
the mass loss to occur on the asymptotic giant branch.
\begin{table}
  \caption{Wind Parameters for the Various MESA Models.}
  \centering
\label{tab:windparam}
\begin{tabular}{ccccccc}
\hline
$\eta_\mathrm{RGB}$ & $\eta_\mathrm{AGB}$ & $M_\mathrm{HB}$ & $M_\mathrm{WD}$  & $\Delta t_\mathrm{T2T}$ & $\Delta t_\mathrm{CHB}$  & $t_\mathrm{WD}$ \\
                 &                    &  [$M_\odot$] & [$M_\odot$] & [Myr] & [Myr] & [Gyr] \\
\hline
0.1  & 0.7  & 0.86 & 0.54 & 108  & 89.2 & 10.8 \\
0.2  & 0.6  & 0.82 & 0.54 & 109  & 89.7 & 10.8 \\
0.3  & 0.5  & 0.78 & 0.53 & 110  & 90.4 & 10.8 \\
0.46 & 0.46 & 0.70 & 0.53 & 111  & 91.0 & 10.9
\end{tabular}
\end{table}

The procedure for the PARSEC-COLIBRI models was similar, but for the
mass-loss descriptions.  A first set of models was computed adopting
the Reimers law on the RGB, in combination with the
\citet{SchroederCuntz_05} formula for the TP-AGB, as modified by
\citet{Rosenfield_etal14}:
\begin{equation}
\begin{split}
{\dot M}_ {\rm mSC} =   10^{-12} \eta_{\rm mSC}
 \frac{L}{L_\odot} \frac{R}{R_\odot}
\frac{M_\odot}{M}
\left(\frac{T_{\rm eff}}{4000\, {\rm K}}\right)^{8.9}\\
\left(1+\frac{g_{\odot}}{4300 g}\right) \,\,\,\,\,\,\,\,\,\,\,\,\,[{M_\odot}{\mathrm{yr}}^{-1}]
\label{eq_mSC}
\end{split}
\end{equation}
where $g=G M /R^2$ is the surface gravity, and $\eta_{\rm mSC}$ is a
free efficiency parameter.  Again for each value of $\eta_\mathrm{RGB}
=0.1, 0.2, 0.3, 0.4$, the parameter $\eta_{\rm mSC}$ of the AGB mass
loss was tuned so as to obtain a final C-O core with mass of $\simeq
0.53-0.54\, M_{\odot}$.  We also found that the tip of the TP-AGB in
PARSEC-COLIBRI models was consistent with the observed $M_V$ at the
tip of the TP-AGB in 47~Tuc \citep{2005A&A...441.1117L}.

We recall that Eq.~(\ref{eq_mSC}) was proposed by
\citet{Rosenfield_etal14} with the purpose of reproducing the star
counts and luminosity functions of TP-AGB stars detected in a sample
of nearby metal-poor galaxies from the ACS Nearby Galaxy Survey
Treasury \citep[ANGST;][]{Dalcanton2009}, which do not present signs
of recent star formation.  The working scenario is that at lower
luminosities on the AGB winds are not driven by dust opacity, rather
they should be linked to the mechanical flux produced in highly
turbulent chromospheres \citep{CranmerSaar_2011}.  Adopting
$\eta_\mathrm{RGB}=0.2$ for RGB stars, the ANGST data were best
reproduced assuming that on the TP-AGB low-metallicity low-mass stars
experience more mass loss than predicted by the classical Reimers law
(with $\eta_\mathrm{RGB}=0.3-0.4$).  Good agreement with observations
was obtained by setting the efficiency parameter $\eta_{\rm mSC}=0.4$
in the modified \citet{SchroederCuntz_05} formula.  The calibration
also sets constraints on the TP-AGB lifetimes for $\mathrm{[Fe/H]}\la
-0.9$.  They should be $\sim 0.5$~Myr for lower mass stars ($M \la 1\,
M_{\odot}$), corresponding to final masses of $0.52-0.54\, M_{\odot}$.

A second test calculation was performed with the prescriptions
recently proposed by \citet{Origlia_etal14} to describe mass loss on
both the RGB and the AGB of Galactic globular clusters.  This
formulation was derived relating the mid-IR excess exhibited by a
fraction of RGB stars to the mass-loss rate, through a scaling
relation that involves a few parameters for the dust properties.  We
set the parameters equal to the reference values suggested by the
authors for 47 Tucanae, namely: expansion velocity
$\upsilon_\mathrm{exp}= 10$ km s$^{-1}$, gas-to-dust ratio
$\delta=200$, and grain density $\rho=3$ g cm$^{-3}$.  Another
quantity to be specified is the fraction $f_{\rm on}$ of dusty RGB
stars, which may vary with the bolometric magnitude $M_{\rm bol}$.
Its value was derived from Fig.~4 of \citet{Origlia_etal14} paper.
Assuming $\mathrm{[Fe/H]}=-0.7$ for 47 Tucanae, we got $f_{\rm
  on}=0.098$ for $-1.5 \le M_{\rm bol} \le -0.6$ and $f_{\rm
  on}=0.222$ for $M_{\rm bol} < -1.5$ on the RGB; $f_{\rm on}=0.292$
for $M_{\rm bol} < -1.5$ on the AGB. No mass loss was considered for
$M_{\rm bol} > -0.6$.

A third test calculation was carried out using another semi-empirical
mass loss relation based on the measured infrared excess. With the aid
of dust radiative transfer calculations that best fit the observed
spectral energy distributions of a sample of field RGB stars, with
accurate parallaxes, \citet{Groenewegen_12} derived the formula
\begin{equation}
{\dot M}_R =   4 \times 10^{-14}
\left(\frac{L}{L_\odot} \frac{R}{R_\odot} \frac{M_\odot}{M}\right)^{0.90}
\,\,\,\,\,\,\,\,\,\,\,\,\,[{M_\odot}{\mathrm{yr}}^{-1}]\,.
\label{eq:groen}
\end{equation}
We note that there is no adjustable parameter here, and that the
reference dust parameters used by \citet{Groenewegen_12} were
$\delta=200$, and $\upsilon_\mathrm{exp}= 10$ km s$^{-1}$, which are
exactly the same as the ones adopted in the \citet{Origlia_etal14}
formulation.  Instead, as we will see below, the predictions in terms
of RGB mass loss are significantly different!  Later on the AGB mass
loss was described with the modified \citet{SchroederCuntz_05} formula
given by Eq.~(\ref{eq_mSC}).  An efficiency parameter $\eta_{\rm
  mSC}=0.8$ was found to be the suitable choice to obtain a final mass
of $\simeq 0.54 \, M_{\odot}$.

Table~\ref{tab:windparam-colibri} outlines the results of the
PARSEC-COLIBRI models.  The quantitative trends are quite similar to
those obtained with the MESA code. Small differences in the mass on
the horizontal branch for the same $\eta_\mathrm{RGB}$ can be easily
explained by the details of the model.  Concerning the model with the
\citet{Origlia_etal14} mass loss, we just note that it yields a final
mass of $0.57\, M_{\odot}$, that is larger than our reference value of
$0.53\, M_{\odot}$. In fact, with the \citet{Origlia_etal14} formula
for mass loss the TP-AGB star is predicted to experience 9 thermal
pulses before leaving the AGB, while in all other COLIBRI models the
total number of thermal pulses is $\simeq 3-4$. The corresponding
lifetime is therefore longer, $\simeq 2.1$ Myr, compared to $\simeq
0.5-0.9$ Myr for the set of TP-AGB calculations made with the modified
\citet{SchroederCuntz_05} relation.

Even though the method to derive the mass-loss rates for RGB stars is
intrinsically similar to that adopted by \citet{Origlia_etal14}, {\em
  i.e.} dust radiative transfer calculations that best fit the
spectral infrared excess, the predictions for RGB mass loss obtained
with the semi-empirical relation of \citet{Groenewegen_12} are
completely different.
\begin{table}
\caption{Wind Parameters for the Various PARSEC/COLIBRI Models.}
\label{tab:windparam-colibri}
\centering
\begin{tabular}{ccccccc}
\hline
$\eta_\mathrm{RGB}$ & $\eta_\mathrm{mSC}$ & $M_\mathrm{HB}$ & $M_\mathrm{WD}$  & $\Delta t_\mathrm{T2T}$  & $\Delta t_\mathrm{CHB}$   &  $t_\mathrm{WD}$ \\
                 &                    &  [$M_\odot$] & [$M_\odot$] & [Myr] & [Myr] & [Gyr] \\
\hline
0.1  & 0.7  & 0.84 & 0.54 & 107 & 94.5 & 10.98 \\
0.2  & 0.5  & 0.79 & 0.54 & 108 & 94.9 & 10.98 \\
0.3  & 0.3  & 0.72 & 0.53 & 111 & 97.9 & 10.98 \\
0.4  & 0.13 & 0.65 & 0.53 & 109 & 95.4 & 10.98 \\
\hline
Or14 & Or14 & 0.69 & 0.57& 111 & 97.9 & 10.98 \\
Gr12 & 0.8 & 0.88 & 0.54& 108 & 96.0 & 10.98 \\
\hline
\multicolumn{2}{l}{References.} &
\multicolumn{4}{l}{Or14:\citet[][]{Origlia_etal14}}\\
\multicolumn{2}{l}{} & \multicolumn{4}{l}{Gr12:\citet[][]{Groenewegen_12}} \\
\end{tabular}
\end{table}

\section{The Epoch of Mass Loss}
\label{sec:epoch-mass-loss}

Through an analysis of the distribution of magnitudes and positions of
the young white dwarfs in the core of 47~Tucanae, \citet{Heyl14diff}
measured the rate of diffusion due to relaxation in the cluster. The
right panel of Fig.~\ref{fig:CMD} depicts the radial distribution of
the brighter and fainter white dwarfs.  Both groups of white dwarfs
have the nearly same mass because they formed from stars of similar mass.
The mass of the turn-off in 47~Tucanae varies by about 0.3\% over
100~Myr, and the mass of the resulting white dwarf would vary by
0.05\% over this same period.  However, the fainter ones were formed
earlier, so they have been diffusing through the cluster for a longer
time.  We have also added the radial distribution of the stars on the
upper main sequence.  This distribution is only slightly more
concentrated than the young white dwarfs, indicating that there has
been very little time for the young white dwarfs to have diffused
through the cluster since their progenitors lost mass.  Furthermore,
the distribution of the upper-main-sequence stars is nearly identical
to that of the red-giant branch stars even when we focus on
seven-hundred stars near the tip of the red-giant branch.  Using the
white-dwarf formation rate from \citet{Heyl14diff} and the PARSEC
models depicted in Fig.~\ref{fig:CMD}, this corresponds to the last
hundred million years along the red-giant branch, indicating that
little mass loss has occurred up to about 100~Myr before the tip of
the RGB. In particular one can pose the question how much diffusion
has occurred between the upper main-sequence or equivalently the RGB
and the young white dwarfs.

\begin{figure}
  \includegraphics[width=\columnwidth]{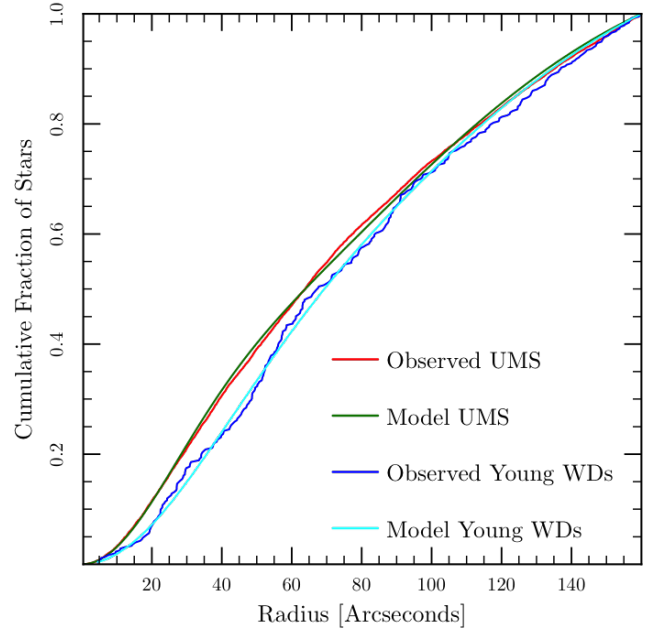}
  \caption{ Radial distribution of the upper-main sequence stars as
    depicted in Fig.~\ref{fig:CMD} and the white dwarfs younger than
    50~Myr. There are 367 young white dwarfs, and their median age is
    18.9~Myr since the tip of the AGB. The diffusion time between the
    two model distributions is 21.9~Myr.}
 \label{fig:raddist}
\end{figure}

We can play the diffusion back in time to what is presumably the
initial radial distribution of the stars, that of stars on the upper
main sequence or the giants as depicted in Fig.~\ref{fig:CMD}.  We
assume a cooling curve for the white dwarfs and the best-fitting
two-Gaussian model for the distribution of the white dwarfs as they
diffuse; all that we vary is the time for main-sequence radial
distribution to diffuse and to appear like the distribution of white
dwarfs younger than 50~Myr.  We fit for the best-fitting timescales by
performing a Kolmogorov-Smirnov test between the diffusion model and
the distribution of stars.  Fig.~\ref{fig:raddist} depicts the best-fitting
diffusion models along with the observed distributions.
If we assume a cooling curve for the white dwarfs as described in
\S~\ref{sec:stell-evol-models} and a two-Gaussian model for the
diffusion, we find that the mass-loss event coincides with the tip of
the AGB, and the mass loss greater than 0.2 solar masses earlier than
20~Myr before the tip of the AGB can be excluded with ninety-percent
confidence.  A major mass-loss event on the RGB (greater than
0.2$M_\odot$ can be excluded at the greater than the four-sigma level
($p<10^{-8}$) from the diffusion of the youngest white dwarfs alone.

\begin{figure}
\includegraphics[width=\columnwidth]{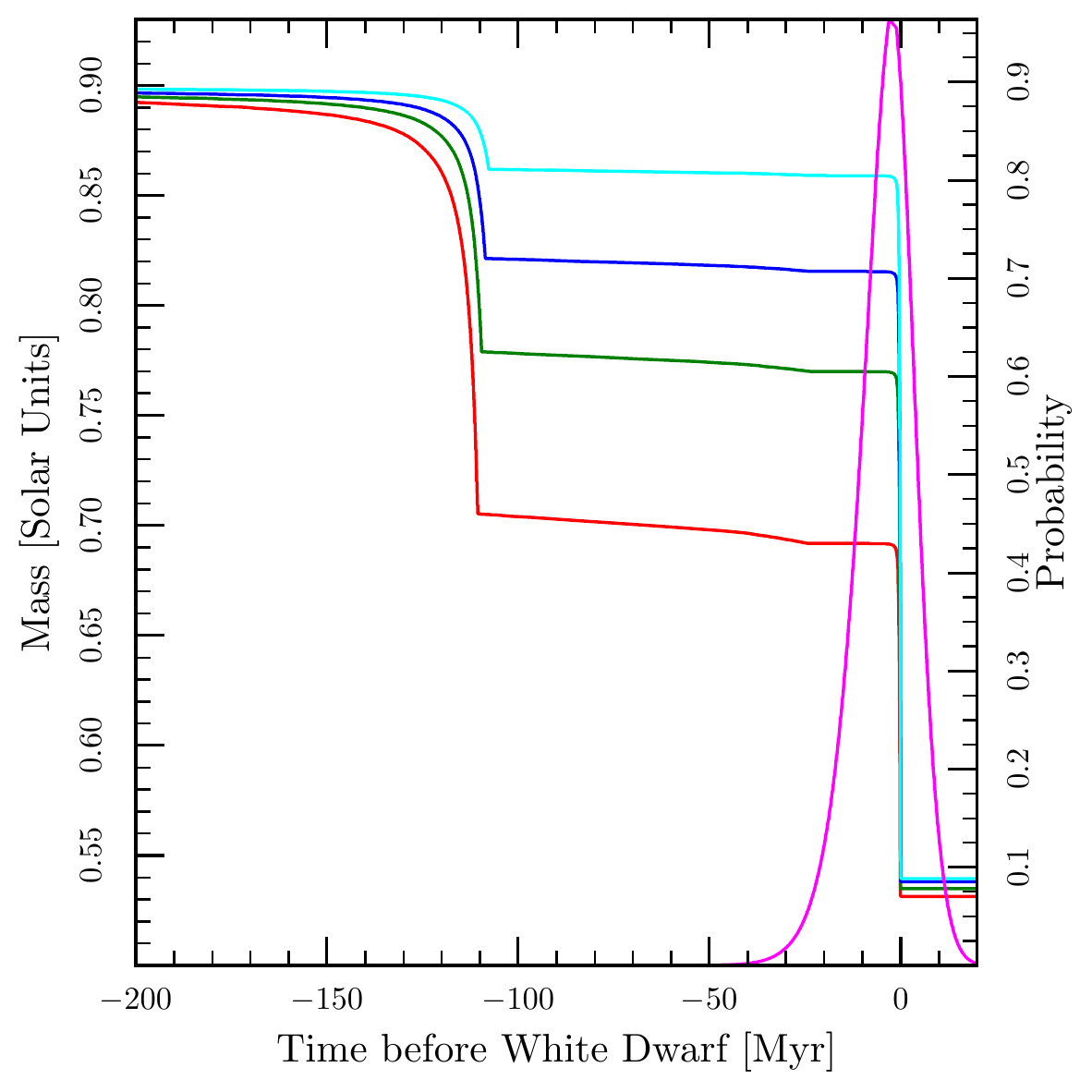}
\caption{Mass of a star that will become a modern-day white dwarf in
  47 Tucanae as a function of time before its peak luminosity for the
  MESA models.  The first epoch of mass loss in the model is as a red
  giant and the second is as an asymptotic giant star.  The lowermost
  curve traces the mass loss for
  $\eta_\mathrm{RGB}=\eta_\mathrm{AGB}=0.46$ and the upper curves
  trace the mass loss for $\eta_\mathrm{RGB}=0.1, 0.2$ and $0.3$ from
  top to bottom with the other quantities given in
  Table~\ref{tab:windparam}. The magenta curve depicts the KS
  probability of the epoch of mass loss obtaining calculating the
  diffusion of the UMS stars over a given time interval and comparing
  the resulting radial distribution with that of the young white
  dwarfs (median age 18.9~Myr) --- the best fitting time for the mass
  loss to have occurred coincides with the TP-AGB.}
\label{fig:time-mass}
\end{figure}

\begin{figure}
\includegraphics[width=\columnwidth]{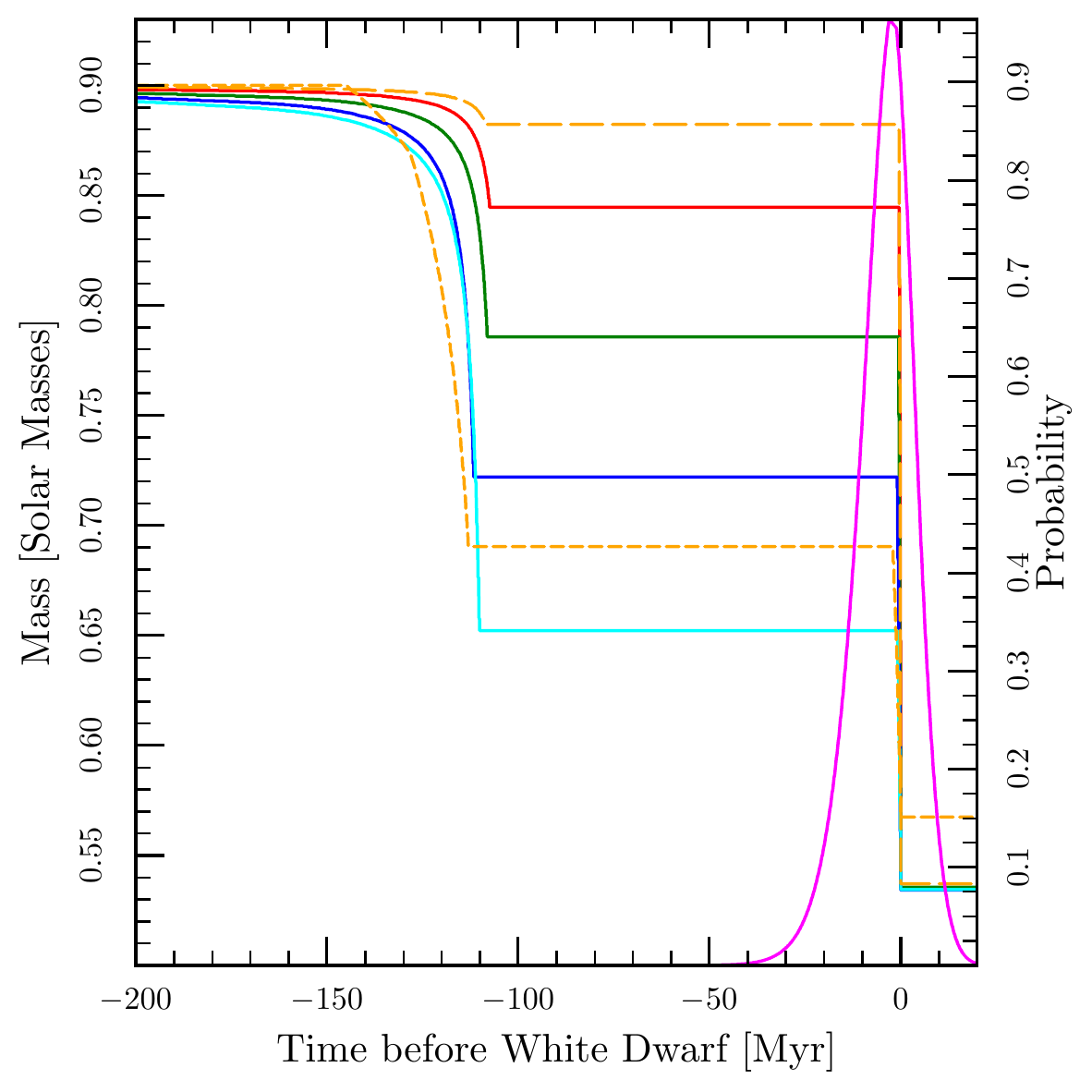}
\caption{Mass of a star that will become a modern-day white dwarf in
  47 Tucanae as a function of time before its peak luminosity for the
  COLIBRI models.  From top to bottom the four curves (solid lines)
  trace the mass loss for $\eta_\mathrm{RGB}=0.1, 0.2, 0.3$, and $0.4$
  with the other quantities given in
  Table~\ref{tab:windparam-colibri}.  The short-dashed line
  corresponds to the \citet{Origlia_etal14} formulations for mass
  loss, while the uppermost long-dashed line refers to the mass-loss
  rates predicted by \citet{Groenewegen_12} for the RGB and with the
  modified \citet{SchroederCuntz_05} relation for the AGB.  Note that
  in both cases the RGB mass loss is derived from the results of the
  semi-empirical infrared-excess method, with quite different
  results.}
\label{fig:time-mass-colibri}
\end{figure}

Fig.~\ref{fig:time-mass} and~\ref{fig:time-mass-colibri} also depict
the Kolmogorov-Smirnov probability obtained by evaluating the
diffusion time-scale required to go from the best-fitting model for
the UMS stars to the best-fitting model for a sample of young WDs
(median age of 18.9~Myr since the tip of the AGB) as depicted in
Fig.~\ref{fig:raddist}.  This brings the epoch of major mass loss
practically coincident with the TP-AGB.  This probability assumes that
the theoretical diffusion model is fixed; in particular, it does not
include the approximately ten-percent uncertainty in the diffusion
timescale.  This would shift the peak of the probability about 2~Myr
in either direction.  The statistical standard deviation in
determining the median age of the white-dwarf sample is 1.2~Myr.  The
theoretical cooling curve itself also yields an additional uncertainty
in the diffusion timescale of about 4~Myr.  This is obtained by
comparing of the theoretical cooling curve with the empirical cooling
curve \citep{2012ApJ...760...78G}.  Combining these uncertainties in
quadrature yields the result that the mass-loss should have taken
place $3.0 \pm 8.1$~Myr before the tip of the AGB.

The indications from the diffusion modelling show that the bulk of the
mass loss is likely to have occurred while the white dwarf progenitors
ascended the asymptotic giant branch.  The distribution of the young
white dwarfs closely resembles that of the upper main sequence stars
because they have not had time to relax to their new mass.  It is
unclear whether all of the mass loss occurs on the asymptotic giant
branch or just the bulk of it.  From these dynamical arguments it is
likely that the mass-loss evolution calculated from the MESA model
with equal values of $\eta$ depicted by the lowermost curve in
Fig.~\ref{fig:time-mass} is not correct.  In this case during the
first mass loss event, the mass of the progenitor decreases by 0.20
solar masses on the red-giant branch and 0.17 solar masses on the
asymptotic giant branch.  Similar considerations apply to the
PARSEC-COLIBRI models as depicted in Fig.~\ref{fig:time-mass-colibri}
with $\eta_\mathrm{RGB}=0.4$. In this case the mass lost on the
red-giant branch, 0.25 solar masses, is even larger, by a factor of
two, than the mass expelled on the asymptotic giant branch, 0.12 solar
masses.  The model computed with the \citet{Origlia_etal14} formalism
predicts similar amounts of ejected masses, with the notable
difference that the final mass and the TP-AGB lifetime are larger.
Two of the diffusion models point to the AGB as the location of the
mass loss, so it is likely that even more than two-thirds of the mass
loss occurs then.

The conclusion is that in 47 Tuc the mass of the horizontal-branch
should be not much different from that the main-sequence stars.
\citet{2012MNRAS.419.2077M} argued from asteroseismic observations of
red giant and red clump stars in NGC~6791 that at least in this
metal-rich cluster little mass is lost on the red-giant branch, less
than a tenth of a solar mass.  If we use the value of $\eta=0.2$ as
suggested by \citet{2012MNRAS.419.2077M} the mass of the star
decreases more on the AGB as traced by the second curve from the top
in Fig.~\ref{fig:time-mass}.  Here the star only loses 0.08 solar
masses on the red-giant branch and 0.28 solar masses on the AGB in
better accordance with the diffusion models.  The predictions of mass
loss with the PARSEC-COLIBRI model are of 0.11 and 0.25 solar masses,
respectively for $\eta_\mathrm{RGB}=0.2$ and $\eta_\mathrm{mSC}=0.5$.
Further support to the findings of our diffusion models comes
from the RGB mass loss rates derived by \citet{Groenewegen_12} from
the measured infrared excess in a sample of field RGB stars.  In fact,
adopting his proposed formulation we obtain that only $\approx 0.02
\,M_{\odot}$ is expelled on the RGB, so that the star reaches the AGB
with a mass that is in practice the same it had on the main
sequence. In this case $\approx 0.34 \,M_{\odot}$ is lost on the
AGB. We emphasize that the agreement was obtained with the original
\citet{Groenewegen_12} relation for the RGB mass loss rates, without
introducing any efficiency parameter. Moreover, as discussed by
\citet{Groenewegen_12}, the mass-loss rates derived in his study are
consistent with the chromospheric estimates for RGB stars.
It also appears that, surprisingly, the winds in metal-rich stars may
operate similarly to those in more metal-poor stars such as in 47~Tuc.

The radial distribution of the horizontal-branch stars themselves may
help to constrain their masses.  The core-helium burning phase on the
horizontal branch lasts for about 100 Myr and stars will linger in the
region of CMD that we define as the horizontal branch for about
80~Myr; this is more than a relaxation time of 30~Myr
\citep{Heyl14diff}, so if there is a significant amount of mass loss
before the horizontal branch, one would expect the horizontal-branch
stars to be less concentrated than the upper main
sequence. Fig.~\ref{fig:CMD} shows that the horizontal branch of the
evolved blue-straggler stars could contaminate the HB region of the
CMD due to saturation.  To remove the saturated stars from the sample
we have imposed a mild cut on the uncertainty in the magnitude
estimates.  Fig.~\ref{fig:CMD} also depicts the radial distribution of
the giants and horizontal-branch stars.  The giant stars have a
similar distribution to the upper-main sequence, and so do the
horizontal-branch stars.
If the horizontal-branch stars and the upper-main sequence stars were
drawn from the same distribution, one would expect to find a deviation
in the cumulative distribution as large as observed more than one
third of the time (using the Kolmogorov-Smirnov test).  Furthermore
the number of horizontal branch stars after correcting for
incompleteness (563) is in loose agreement with the theoretical duration of
the horizontal branch of 80~Myr and the birth rate of white dwarfs in
the field of about 7~Myr$^{-1}$ \citep{Heyl14diff}.

Fig.~\ref{fig:CMD} also depicts the radial distribution of
main-sequence stars whose mass is about $0.65\mathrm{M}_\odot$ as
determined from the PARSEC isochrone (in lavender); the
distribution nearly coincides with the faint white dwarfs (in green).
If the mass of the horizontal branch were $0.65\mathrm{M}_\odot$, this
is the radial distribution to which the HB branch stars would
ultimately relax.  Of course, this would take about 100~Myr to reach
completion, and the average age of a horizontal branch star is only
40~Myr but the bulk of the relaxation would occur within 40~Myr.  The
null hypothesis that the horizontal branch and the main-sequence stars
of a mass of about $0.65\mathrm{M}_\odot$ are drawn from the same
radial distribution can be rejected by a Kolmogorov-Smirnov test at
the six-sigma level ($p\approx 3\times 10^{-17}$).

By comparing the distribution of upper-main-sequence stars, giant
stars, horizontal branch stars, young and old white dwarfs, we find
that it is most likely that the stars that are currently evolving to
form white dwarfs lose more mass as asymptotic giant branch stars than
as red-giant-branch stars.  In the context of the Reimers and
Bl\"ocker or modified Schr\"oder \& Cuntz models for wind mass loss,
parameters such as $\eta_\mathrm{RGB}\approx 0.1$ and
$\eta_\mathrm{AGB} \approx 0.7$ best account for the observed radial
distributions of the stars.
Conversely,  all wind descriptions that predict the bulk of mass loss on the RGB
conflict with the indications presented in this work.

\section{Conclusions}
\label{sec:conclusions}

We must emphasize that the results presented here proceed from two
independent observations.  The first is an inference of the time-scale
of the major mass loss which rules out the RGB in favor of the AGB
phase.  This conclusion is nearly independent of the assumed distance
to 47~Tucanae and the models of the horizontal-branch evolution.  The
time scale that we derive rests on three independent arguments: the
dynamical relaxation time from theoretical considerations, the
white-dwarf cooling timescale and the duration of the red-giant-branch
evolution that can be used to estimate the ages of the white dwarfs
without reliance on white-dwarf cooling models
\citep{Heyl14diff} --- we count the number of
red-giant stars in the CMD and determine the white-dwarf birthrate
from the theoretical evolutionary timescale through this portion of
the CMD \citep[see][for further details]{2012ApJ...760...78G}.
All three agree and point to the AGB as the origin of the mass loss.

Furthermore, even a large relative error in these timescales would
only result in a slight change in the timescale of the mass loss
relative to the duration of the horizontal branch because we are
measuring the time elapsed between the mass-loss event and the
appearance of a young $\sim 20$~Myr white dwarf.  We find this to be
about 20~Myr, so our estimate of the diffusion timescale would have to
be underestimated by a factor of five to place the bulk of the mass
loss on the red-giant branch.  In this case it would be difficult to
account for the diffusion between the bright and faint white dwarfs
unless the white-dwarf cooling timescale were also underestimated by a
factor of five, and the white-dwarf birthrate were overestimated by a
factor of five as well.  Because the white dwarfs are produced through
stellar evolution, a revision of the white-dwarf birth rate would
require a revision of the timescales for the entire post-main-sequence
stellar evolution to achieve consistency with the numbers of stars
observed in the CMD of 47~Tuc.  In particular this would also increase
the duration of the horizontal branch by about a factor of five as
well, leading to further theoretical and observational
inconsistencies.

The second line of evidence rests on the observed radial distribution
of horizontal branch stars that strongly resembles the distribution of
the upper-main sequence stars.  The horizontal branch lasts long
enough to suffer from diffusion if a significant amount of mass loss
occurs before it.  By using a mild cut in magnitude error we have
eliminated the most saturated stars from our horizontal-branch sample.
As shown in Fig.~\ref{fig:CMD}, these saturated stars are most likely the
descendents of the blue-straggler stars.  The number of horizontal
branch stars in the field is also in accord with theoretical models
and rate of formation of white dwarfs in the field.

Our conclusions face two possible difficulties: the presence of
multiple populations in 47~Tuc
\citep{2009ApJ...697L..58A,2012ApJ...744...58M} and the conclusion
from the current state of the art of horizontal-branch modelling that
more mass loss is required to account for the observed horizontal
branch \citep{2010MNRAS.408..999D}.  \citet{2012ApJ...744...58M} found
that the second generation dominates the population most strongly in
the center of the cluster and the ratio of the two populations is
constant with radius within the error bars where our observations
focus.  Although \citet{2013ApJ...771L..15R} found dynamical
signatures and a radial gradient in the two populations, those
conclusions were based on observations far from the cluster core.  In
that outer field the relaxation time is much longer than in the core,
so these initial differences have not yet been erased. The short
dynamical time in the core compared to the age of the cluster ensures
that dynamically the two populations behave similarly; furthermore,
the contribution of the first population is most modest in the core.

On the second front, recent synthetic horizontal branch models
\citep[e.g.][]{2010MNRAS.408..999D} argue that the mass of the
horizontal branch stars is significantly less than that of the mass
sequence stars by about $0.27 M_\odot$.
However, a more firm conclusion on theoretical grounds may come only
considering the simultaneous matching of the whole CMD of the cluster
and exploring the possible degeneracy between various parameters
(i.e. helium content, metal mixture, and mass-loss efficiency).
A detailed
comparison both along the main sequence and throughout the post-main
sequence evolution (especially the HB and AGB) with the data in all
available bands including the ultraviolet would test the models and
either support or contradict the conclusions drawn here from dynamical
evidence.

A key piece of evidence that could bolster these arguments would be a
direct measurement of the masses of the red giant and
horizontal-branch stars in 47~Tucunae, perhaps through
asteroseismology as \citet{2012MNRAS.419.2077M} did in NGC~6791 or by
radial velocity measurements of binaries that include stars in these
evolutionary stages. In any case the findings of this paper on
47~Tucanae are in clear contrast with other independent studies that
indicate the RGB as the phase of major mass loss
\citep{2015MNRAS.448..502M,Origlia_etal14}.  At the same time, they
will also set a strong challenge to stellar evolution models,
especially with regard to the detailed reproduction of the morphology
of the CMD (in particular the HB) of this cluster which is known to
host two stellar populations with peculiar chemical mixtures and
slightly different helium abundances \citep{Milone_etal12}.

\medskip

The authors would like to thank Francesco Ferraro for providing the
star catalogue for the \citet{2006ApJ...652L.121B} paper.  We also
thank L\'eo Girardi, Alessandro Bressan, and Josefina Montalban for
useful discussions.  This research is based on NASA/ESA Hubble Space
Telescope observations obtained at the Space Telescope Science
Institute, which is operated by the Association of Universities for
Research in Astronomy Inc. under NASA contract NAS5-26555. These
observations are associated with proposal GO-12971 (PI: Richer).  This
work was supported by NASA/HST grants GO-12971, the Natural Sciences
and Engineering Research Council of Canada, the Canadian Foundation
for Innovation and the British Columbia Knowledge Development
Fund. This project was supported by the National Science Foundation
(NSF) through grant AST-1211719. It has made used of the NASA ADS and
arXiv.org.  P.M. acknowledges support from the ERC Consolidator Grant
funding scheme ({\em project STARKEY}, G.A. n. 615604), and from the
University of Padova, ({\em Progetto di Ateneo 2012}, ID:
CPDA125588/12).

\bibliography{evolution}
\bibliographystyle{apj}

\end{document}